\documentstyle[psfig,12pt]{article}
\newcommand{\etal}{{\sl et al.}}

\begin{document}
\noindent
{\Large\bf Highly polarized structures in the near-nuclear regions 
of Cygnus~A: intrinsic anisotropy within the cones?}
\vglue 0.5cm\noindent
{\bf C.N. Tadhunter$^{1}$, W. Sparks$^{2}$, D.J. Axon$^{3}$, L. Bergeron$^{2}$,
N.J. Jackson$^{4}$, C. Packham$^{5}$, J.H. Hough$^{3}$, A. Robinson$^{3}$, 
S. Young$^{3}$}
\vglue 0.5cm\noindent
{\small $^{1}$Department of Physics, University of Sheffield, Sheffield S3 7RH, UK \\
$^{2}$Space Telescope Science Institute, 3700 San Martin Drive, Baltimore,
MD 21218, USA \\
$^{3}$Division of Physics and Astronomy, Department of Physical Sciences,
University of Hertfordshire, College Lane, Hatfield, Herts AL10 9AB, UK \\
$^{4}$Nuffield Radio Astronomy Laboratory, Jodrell Bank, University
of Manchester, UK \\
$^{5}$Isaac Newton Group, Sea Level Office, Apartado de Correos,
321, 38780 Santa Cruz de La Palma, Islas Canarias, Spain}
\vglue 0.5cm\noindent
{\bf Abstract.}
We present near-IR imaging polarimetry observations of the nucleus of Cygnus~A,
taken with the NICMOS camera of the HST at a wavelength of 2.0$\mu$m.
These maps reveal a highly collimated region of polarized emission
straddling the nucleus and extending to a radius of 1.2 arcseconds. Remarkably,
this feature coincides with one, but only one, limb of the edge-brightened
bicone structure seen in the total intensity image.  The
high degree ($P_k \sim 25$\%) and orientation of the extended polarization feature are
consistent with a scattering origin. Most plausibly,
the detection of polarization along only one limb of the bicone is a consequence
of {\it intrinsic anisotropy} of the near-IR continuum within the radiation cones, with the direction of maximum intensity of the near-IR radiation field
significantly displaced from the direction of the radio axis.
The unresolved nuclear core source
is also highly polarized ($P_k > 28$\%), with a position 
angle close to the perpendicular
to the radio axis. Given that this high degree
of nuclear polarization can only be explained
in terms of dichroic extinction if the dichroic mechanism is
unusually efficient in Cygnus~A, it is more likely that the nuclear
polarization is due to the scattering of nuclear light in an unresolved
scattering region close to the AGN. In this case, the flux of the core
source in the K-band is dominated by scattered rather than transmitted quasar light, and
previous extinction estimates based on K-band photometry of the core 
substantially underestimate the true nuclear extinction.

\section{Introduction}

As yet, little is known about the structure of the inner regions of
powerful radio galaxies and the impact of the activity on 
circumnuclear regions.
The high resolution and sensitivity afforded by
imaging observations using the Hubble Space Telescope (HST) has revealed
a wealth of complex structures in such objects, with dust lanes, jets
and large regions of scattered emission from hidden nuclei. One of
the key targets in studies which aim to elucidate the interplay between active
nuclei and their host galaxies, and between galactic activity of
different types, is the archetypal powerful radio galaxy Cygnus~A
(see Carilli \& Barthel 1996 for a review).
%As part of a detailed ongoing study
%of this object, 
%here we present
%infrared imaging polarization
%observations centered at $\approx 2.0\mu$m,
%obtained for Cygnus~A using the
%Near Infrared Camera and MultiObject Spectrometer (NICMOS) on-board HST.

HST infrared imaging observations of Cygnus~A by Tadhunter \etal\ (1999) 
revealed an edge-brightened 
bi-conical 
structure centred on the nuclear point source, which is
strikingly similar to those observed around young stellar objects. The 
edge-brightening of this structure provides evidence that the bicone 
is defined as much by outflows in the nuclear regions 
as by the polar diagram of the illuminating quasar radiation field.
The HST observations also show an unresolved nuclear source at 2.0 and 2.25$\mu$m. However, from the imaging observations alone it is unclear
whether this unresolved source represents the highly extinguished quasar nucleus
seen directly through the obscuring torus, or emission from a 
less-highly-extinguished extended region around the nucleus. 
%A further
%implication of the NICMOS images is that not all of the anisotropy in the %nuclear
%radiation field is caused by extinction on a scale $<$100 pc in the torus; some %of the 
%anisotropy must 
%be generated by absorption and 
%scattering in the  dust lane on a 1kpc scale.

Near-IR polarization observations have the potential to remove the uncertainties surrounding
the nature of the unresolved  nuclear sources and, in addition, to provide 
further important
information about the obscuration and anisotropy in the near-nuclear regions
of powerful radio galaxies. Previous ground-based polarimetric
observations of Cygnus~A by 
Packham \etal\ (1998) demonstrate that the nuclear regions are highly polarized
in the K-band, with a measured polarization of $P_k \sim 4$\% for
a 1 arcsecond diameter aperture centred on the compact IR nucleus. However, the 
resolution of the ground-based observations is insufficient to resolve the
structures and determine the polarization mechanism unambiguously.
In this letter we present new diffraction-limited infrared imaging 
polarimetry observations
of Cygnus~A made with the Near Infrared Imaging Camera and Multi-Object
Spectrometer (NICMOS: MacKenty \etal\ 1997) on the HST. 
These observations resolve the polarized structures, and raise new
questions about the nature of the anisotropy in the near-nuclear regions
of this key source.

\section{Observations and data reduction}

NICMOS Camera 2 `long wavelength' infrared imaging polarization observations were taken December 1997 and August 1998, giving a pixel scale of 0.075 arcseconds and
a total field of 19.4$\times$19.4 arcseconds. 
The NICMOS polarizers are self-contained spectral elements, with the three long
wavelength polarizers effective from $\lambda \approx 1.9$---$2.1\mu$m, resulting
in an effective central wavelength of $\approx 2.0\mu$m
(MacKenty \etal\ 1997).
The polarizers are oriented at approximately 60$^{\circ}$ intervals and have
characteristics as presented by Hines (1998).
Each exposure
consisted of a number of non-destructive reads of the detector which were
optimally combined in the reduction software to remove cosmic rays.
Regular chops were made to offset fields in order to
facilitate accurate background subtraction. 
The total integration time was 2400 seconds per polarizer.

The reduction of the data used standard IRAF/STSDAS pipeline processing
together with pedestal removal as given by
van de Marel (1998).
The three final, clean polarization images were combined following the
prescription of Sparks \& Axon (1999), using the pipeline produced
variance data. The output data comprised a set of images containing
each of the Stokes parameters $I, Q, U$, their variances and covariances,
a debiassed estimate of the polarization intensity and polarization
degree using the method of Serkowski (1958), and also position angle and
uncertainty estimates on each of those images, as described in detail
in Sparks \& Axon (1999).

The two epochs of observation were acquired with different instrumental
orientation, and were analysed independently to provide a robust check
on our estimated uncertainties and on the possibility of systematic errors.
A variety of spatial resolutions were used, from full 
diffraction-limited NICMOS resolution down to $\approx 1$ arcsec 
by smoothing the
input three images prior to polarization analysis.
Both epochs were fully consistent within the estimated statistical errors,
implying that there are no significant sources of systematic uncertainty.
Measurements of several Galactic stars in the field of Cygnus~A give 2.0$\mu$m
polarizations of $P_{2.0\mu m} < 2.0$\% for 5 pixel diameter apertures. This
can be regarded as an upper limit on the level of systematic error in estimates
of P. The typical statistical uncertainties for the most highly polarized
regions measured in our full resolution images are $\pm$1.5\% for $P$, and $\pm$3$^{\circ}$
for the polarization angle.
In the following, we will present the data for the December
1997 observations only, since these have a higher S/N 
as a consequence of longer exposure
times.

\section{Results}

Figure 1 shows the first epoch polarization results.
As expected, the total intensity image (Stokes $I$) is very similar
to the direct images published previously by Tadhunter \etal\ (1999).
In particular, it shows
an apparent edge-brightened, reasonably symmetric, bi-conical structure centred on the nucleus
with an 
opening
angle of 116 degrees and whose axis is closely aligned 
with the large scale radio
jet.

The image of polarized intensity, however, reveals
intriguing differences compared to total intensity.
The only regions of strongly polarized
emission (apart from the nucleus discussed below) are confined
to a quasi-linear structure running along the NW-SE limb of the
bi-conical structure. This feature shows approximate reflection symmetry
about the nucleus, as opposed to the axial symmetry about the
radio axis in the total
intensity image.
The brightness ratio of the two limbs of the cone to the east
of the nucleus is approximately
2:1 in total intensity image, while in polarized intensity image it is
$>$12:1. Note that the polarization structure
visible in our 2.0$\mu$m image is strikingly different from that of
the optical V- and B-band polarization images (Tadhunter \etal\ 1990, 
Ogle \etal\ 1997), in which the
polarized emission appears uniformly distributed across the
kpc-scale ionization cones and shows no clear preference for the
NW-SE limb of the bi-cone. 

Typical {\it measured} degrees of polarization
are in excess of $10$\%\, up to a maximum of $\approx$25\%
in the polarized region to the south east of the nucleus. However, 
these measures underestimate the true degree of {\it intrinsic}
polarization in the extended structures, because starlight from the host
galaxy makes a substantial contribution to the total flux. For example, using
the azimuthal intensity profile measured in annulus with inner radius 4 pixels
and outer radius 12 pixels,
we estimate that the diffuse starlight from the host galaxy contributes
50 - 70\% of the total flux in the south east arm of the bicone. Assuming that
the starlight is unpolarized, the degree of intrinsic polarization in
the south east arm is $P_{2.0\mu m}^{intr} \sim$50 -- 70\%. Such high degrees of measured
and intrinsic IR polarization are unprecedented in observations of active galaxies
in which the synchrotron emitting jets are not observed directly at infrared
wavelengths. 
%
% Is this last sentence true?? 
%

%Particularly the linear feature to the SE of the nucleus, is highly
%polarized at a level of several tens of percent. Thus it is likely that the bi-cone 
%represents  an illuminated structure --- an inner reflection nebulosity. 
%Further support for the idea 
%that
%the bicone is dominated by reflected continuum light at infrared wavelengths is provided by 
%the
%fact that the structure is visible in filters (e.g. F160W) which do not suffer substantial
%emission line contamination. 
%In addition to the reflected nuclear light, it is likely that there
%is a significant contribution to the bi-cone
%emission from Pa$\alpha$ emitted by
%photoionized regions in the illuminated structures (Pa$\alpha$ falls within the bandpass
%of the polarization filters).

\subsection{The nucleus}

The nuclear point source, discussed in detail by Tadhunter \etal\ (1999),
is also highly polarized. In the polarized intensity image the main nuclear
component
appears unresolved ($FWHM = 2.24$ pixels) and its position agrees
with that of the nucleus in the total intensity image to within 0.5 pixels (0.04 arcseconds).
Thus, it appears likely that the bulk of the polarization is associated with the compact
nucleus rather than a more extended region around the nucleus. The core
does, however, show a faint entension to the NW in the polarized intensity image.
This extension is aligned with the larger scale polarization
structures, and its polarization E-vector is close to perpendicular to the radius vector from the nucleus.

From our full resolution polarization images, the measured
degree of polarization at the peak flux of the nuclear point 
source is $P_{2.0\mu m}^{m} \sim 20$\%.
However, spurious polarization can arise because of small mis-alignments between
the polarization images, especially in the nuclear regions where
there are sharp gradients in the light distribution. To guard against such effects
we have smoothed the polarization data using a 5$\times$5 pixel boxcar filter
(0.375$\times$0.375 arcseconds) and re-measured the polarization in the nuclear regions. As expected, the measured degree of polarization in the nucleus in the smoothed image
is less ($P_{2.0\mu m}^{m}\sim10$\%) than in the full resolution image, because of the greater degree of contamination
by unpolarized starlight and extended structures around the nucleus. In order
to determine the intrinsic polarization of the point source it is
necessary to first determine the 
proportion of flux contributed by the point source to the total flux in
the nuclear regions. Experiments involving the
subtraction of a Tiny Tim generated point spread function (Krist \& Hook 1997)
suggest that an upper limit on the 
fractional contribution of the nucleus to the total flux in a 5$\times$5 pixel
pixel box centred on the nucleus is
$f_{nuc} < 35$\%. Thus,
assuming that {\it all} the polarization in the near-nuclear regions is due to the unresolved
compact core, and that the remainder of the light is unpolarized, the intrinsic
polarization of the unresolved core source is $P_{2.0\mu m}^{intr} = P_{2.0\mu m}^{m}/f_{nuc} > 28$\%.

The position angle of polarization E-vector of the 
core source  measured in our full resolution polarization
map ($PA = 201 \pm 3$), is close to perpendicular to the
radio jet axis ($PA = 105 \pm 5$). This is similar to the situation
seen in other AGN, and in particular Cen~A, where, towards longer wavelengths
and smaller apertures, the infrared polarization becomes more and more closely
perpendicular to the radio jet (Bailey \etal\ 1986).

\section{Discussion}

\subsection{The nature of the unresolved core source}

A major motivation for the HST observations was to investigate the nature of the
compact core source and the cause of the relatively large polarization
measured in the core by Packham \etal\ (1998). The explanation
favoured by Packham \etal\ is that the
compact core source represents transmitted quasar light, 
while the high polarization
is due to dichroic absorption by aligned dust grains in the central 
obscuring torus. Because the
dichroic mechanism is relatively inefficient, a high polarization implies a large
extinction: from observations of Galactic stars it is known
that {\it at least} 55 magnitudes of visual extinction
is required to produce a K-band polarization of 28\%
for optimum grain alignment (Jones 1989). More typically, the correlation
between K-band polarization and extinction deduced for Galactic stars
by Jones (1989) implies that an extinction of $A_v \sim 350$ magnitudes
would be required for $P_k = 28$\%. 
For comparison, an upper limit on the K-band extinction in
Cygnus A, estimated by comparing
the 2.25$\mu$m core flux  with mid-IR and X-ray fluxes, is
$A_v < 94$ magnitudes (see Tadhunter
\etal\ 1999 for details).  Thus, the dichroic mechanism is only
feasible if
the efficiency of the mechanism in Cygnus A is greater than it is along most
lines of sight in our Galaxy. Such enhanced efficiency
cannot be entirely ruled out, given that the  Galactic dichroic
polarization involves a randomly oriented magnetic field component, whereas the magnetic fields in the central obscuring regions of AGN may be more coherent.
In this context it is notable that near- and mid-IR polarization measurements of the central regions of the nearby
Seyfert galaxy NGC1068 provide evidence for
a greater dichroic efficiency than predicted by the Jones (1989) correlation,
with $P_k = 5$\% produced by $A_v \sim$20 -- 40 magnitudes (Lumsden et al.
1999). However, even the greater dichroic efficiency deduced for NGC1068   
would not be sufficient to produce the high 
polarization measured in 
the core of Cygnus A if $A_v < 94$ magnitudes.    
%If
%correct, this mechanism would imply 
%the presence of an ordered magnetic field in the near-nuclear
%regions of Cygnus~A, with the field direction
%predominantly in the plane of the obscuring torus.

The efficiency problem might be resolved if the extinction to the core source in
the K-band is higher than the $A_v < 94$ estimated on the basis of comparisons of the K-band flux with the mid-IR and X-ray fluxes. Indeed,
substantially higher extinctions have been deduced for Cygnus A, both from 
modelling the X-ray spectrum of the core
($A_v = 170\pm30$: Ueno \etal\ 1994) and from comparisons between 
hard-X-ray continuum, [OIII] emission line and mid-infrared continuum 
fluxes ($A_v = 143\pm35$: Ward 1996, Simpson 1995). If such high extinctions also apply to the quasar nucleus in the K-band, the low efficiency of
the dichroic mechanism would be less of a problem. However, for any reasonable quasar SED, the
contribution of such a highly obscured quasar nucleus to the flux and the 
polarization of the {\it detected} 2.0$\mu$m core source would 
be negligible (i.e. we would not expect to detect the quasar nucleus directly
in the K-band).
Thus, it is more likely that the relatively low extinction deduced
from the K-band flux measurements reflects contamination of the K-band core by emission from a less-highly-obscured 
region, which is close enough to the central AGN to remain  unresolved
at the resolution of our HST observations. Although it has been proposed that
the contaminating radiation in the K-band may include hot dust emission and/or line emission
from quasar-illuminated regions close to the nucleus
(e.g. Stockton \& Ridgway 1996), such emission would have a low intrinsic polarization, and a large dichroic efficiency would still be required 
in order to produce the polarization of this component by dichroism.

%Given the problems with the dichroic polarization mechanism it is important
%to consider alternative explanations for the high polarization of the core. 
The most plausible alternative to dichroic extinction 
is that the K-band core source
represents scattered- rather than transmitted quasar light. In this case,
the polarization is a consequence of scattering in an unresolved region
close to the illuminating quasar; we do not detect the quasar nucleus
directly in the K-band; and previous extinction estimates based on the K-band
fluxes substantially underestimate the true nuclear extinction. Note
that the presence
of such a scattered component
would resolve the discrepancy between the extinction estimates based on
K-band flux measurements, and those based on fluxes measured at other wavelengths.

Finally, we must also consider the possibility that the core polarization is 
due to synchrotron radiation associated with the pc-scale
jet visible in VLBI radio images (Krichbaum \etal\ 1996). 
Although the {\it integrated} polarization 
of the radio core is small even at high radio frequencies ($P_{22GHz} < 5$\%:
Dreher 1979), we
cannot entirely rule out the possibility that we are observing a highly polarized sub-component of
the jet which suffers a relatively low extinction, or alternatively
that the radio core source as a whole suffers large Faraday depolarization at radio
wavelengths, and would appear more highly polarized at infrared wavelengths.  
Polarization observations at sub-mm wavelengths
will be required to investigate this latter possibility.

%The edge-brightened bi-polar structure is strikingly similar to the structures
%observed around young stellar objects (YSOs: \eg\
%Velusamy \& Langer 1998), which led Tadhunter \etal\ (1999)
%to propose that outflows driven by the central quasar
%hollow out funnels on opposite sides of the nucleus in the kpc-scale disk 
%responsible for the dust lane; the lateral expansion of the outflows
%lead to a density enhancement in the walls of the funnel
%and the structures are
%illuminated by the anisotropic radiation field of the quasar.  
%The opening angles of the illumination cones produced by the pc-scale
%torus surrounding the quasar must be larger than the opening angles of the funnel 
%structures produced by the outflows. 

\subsection{The extended polarization structures}

An intriguing feature of our HST observations is the high degree of polarization measured
along, and only along, the NW-SE limb of the bicone. The
orientation of the polarization measured along the limb is consistent with the scattering
of light from a compact illuminating source in the nucleus,
while the high degree
of polarization is consistent with the edge-brightened bi-cone
geometry of Tadhunter \etal\ (1999), in the
sense that the scattering angle for the edge-brightened region will
be close to the optimal 90$^\circ$ required for maximal polarization.
However, the fact that the polarization is measured along only one
limb is difficult to reconcile with the simplest bicone model
in which the illuminating IR radiation field is azimuthally isotropic, and the
scattering medium is uniformly distributed around the walls of the funnels
hollowed out by the circum-nuclear outflows. In this simplest model
both limbs would be highly polarized in the direction
perpendicular to the radius vector of the source, and this is clearly
inconsistent with the observations.

Our observation require that one or more of the assumptions implicit in the simple model must be relaxed.
In general terms this requires either invoking specific matter distributions
within the cone and/or an anisotropic illumination pattern of the central source 
itself.

Perhaps the simplest way of reconciling the polarization characteristics 
with the bi-cone geometry is to adjust the relative 
importance of scattering and intrinsic emmission with azimuth around the cone,
so that one limb of the cone is dominated by scattered radiation, while the
other is predominatly intrinsic radiation. 
%In order to explain the
%differences between the optical and IR polarization images in this
%way, the scattering material would need be close to optically thick
%to the optical continuum in all 
%directions around the walls of the cones, but only present a substantial
%optical depth to the near-IR continuum in direction corresponding to the
%NW-SE limb of the bi-cone. This wavelength dependence
%would in turn imply that the scattering medium is dust rather than electrons.
Since the band-pass of the 
NICMOS polarizers contains the Paschen alpha line, this provides an obvious potential
source of the diluting radiation for the unpolarized regions.  
%Other possibilites include thermal radiation
%from hot dust and star formation.  
Unfortunately there is no obvious reason why such an asymmetry should exist. 
Furthermore, direct images with the F222M filter show that there is no radical
change in the relative brightness of the two limbs of the eastern cone between
2.0$\mu$m and 2.25$\mu$m. This is an 
argument against  the Paschen alpha model for the intrinsic
emission, since the F222M filter admits no emission lines as
strong as Paschen alpha.
%
%
% 
%
%While star formation might be triggered by an expanding shock 
%wave surrounding the jet, there seem no obvious way to suppress the 
%star-froamtion on the polarized limb,
%and the symmetry in total flux would have to be coincidental. 
% 
% To radiate significantly at 2.0 microns the dust would have to be
% very hot, and it seems unlikely that such high temperatures
% could be produced at such large distances from the nucleus --- most
% models of hot dust place the the dust on a pc rather than kpc scale.
% I also find the star formation idea rather unlikely. So I have 
% left both of these possibilities out (given the lack of space).....Clive
%
%
%
% JIM/TIM How much bigger does the mass density of dust have to be for
% multiple scattering to provide the explanation?
% Isn't multiple scattering a non-starter in the sense that the this
% would imply that the scattering medium is optically thick at IR wavelengths? In this
% case, if it's dust scattering, then the dust lane as a whole would be optically
% thick in the IR and we wouldn't see through to the nuclear regions!  --- Clive
%
%
%Other alternatives are equally unpalatable. For example one could effectively
%depolarise one limb if the optical scattering depth was large enough so that
%multiple scattering was important. Cancelling the polarization by the intriduction of a second 
%source of polarization e.g. dichroic absorption via transmittion through aligned grains requires
%a conspiratorily magnetic field geometry and high optical depths.....Clive
% 

An alternative possibility
is that the NW-SE limb is brighter because the near-IR
radiation field of the illuminating AGN is more intense in that direction (i.e. the illuminating
radiation field is azimuthally anisotropic within the cones). In this
case, the clear difference in structure
between the optical and 
near-IR reflection nebulae suggests that the sources of illumination
at the two wavelengths are different: while the source of the shorter wavelength
continuum must produce a radiation field which is azimuthally isotropic within
cones defined by the obscuring torus, the source of illumination 
at the longer wavelengths is required to display
considerable degree of azimuthal anisotropy. 
The near-IR continuum source must also have a relatively red spectrum, in order
to avoid producing similar structures at optical and infrared wavelengths. 
The near-IR anisotropy might arise in the following ways.
\begin{enumerate}
\item {\bf Beamed radiation from the inner radio jet.} The near-IR continuum 
is emitted by a  
component of the inner synchrotron jet which has a direction
of bulk relativistic motion siginificantly
displaced from the axis of the large-scale radio jet, such that
the radiation is beamed towards the NW-SE limb of the bicone. However, given
the remarkable degree of collimation, and the lack of bending, observed in the Cygnus~A
jet on scales betweem 1pc and 100kpc,
a major difficulty
with this model is that the the jet would have to bend through a large angle on
a scale smaller than $\sim$1pc (the resolution of the VLBI maps), whilst retaining
the rotational symmetry in the jet structure about the nucleus. A further requirement
of this model is that, if
the inner jet is precessing, the precession timescale must be greater than the light
travel time ($\sim$5000 years) across the bicone structure.
\item {\bf Anisotropic hot dust emission.} The near-IR continuum is emitted by hot dust in the inner regions of the galaxy,
with a larger projected area of the emitting region visible from one limb of the bicone
than from the other. For example, if the near-IR radiation is emitted by dust in a warped disk close
to the central AGN --- perhaps an outer part of the accretion disk --- the warp could be
oriented such that the NW-SE limb has an almost face-on view of the emitting region, whereas the
the NE-SW limb has a more oblique view. There is already direct observational evidence for
a warped outer accretion disk in at least one active galaxy
(Miyoshi \etal\ 1995). Given that
this mechanism would produce a relatively mild, broad-beam anisotropy,
an optically thick torus on scales larger than the hot dust emitting region would
still be required in order to
produce the sharp-edges to the illuminated bicone structure.
\end{enumerate} 
%The beaming idea draws some support from
%the fact that some infrared  nebulae associated
%YSOs in our Galaxy
%show a similar morphologies in polarized light, and these features have
%explained in terms of scattering of a highly collimated beams of radiation
%(e.g. the Chamealeon nebula: ref??)
Note that
regardless of how any anisotropy in the IR radiation field might be produced,
such anisotropy would not by itself explain the nature of the unpolarized
emission along the SW-NE limb of the bicone, and the lack of variation
in the brightness
ratio of the two limbs of the eastern cone between 2.0$\mu$m
and 2.25$\mu$m. It may also
be difficult to reconcile the anisotropic illumination model
with the polarization properties of the unresolved
core source: if the core polarization is due to scattering, the orientation
of the core polarization vector implies that a substantial flux of illuminating
photons must escape at large angles to the NW-SE limb of the bicone.

We expect future spectropolarimetry observations to resolve the uncertainties concerning
the origin of the near-IR polarization structures in Cygnus~A.
For example, in the case of the anisotropic illumination mechanisms considered above, the anisotropy
is in the {\it continuum} flux rather than the broad lines
associated with the AGN. Thus,  if this model is correct, 
the broad lines will be relatively weak or
absent in the polarized spectrum of the extended structures.
In contrast, for a non-uniform distribution of scattering material but
isotropic illumination within the cones, the broad lines and continuum will be 
scattered equally, and the equivalent widths of the broad lines in the polarized
spectrum should fall within the
the range measured for steep spectrum radio quasars.

%A more radical explanation might be that the innner morphology is not a true 
%bi-cone but, for example, originates as a projection of a warped disk.
%In this picture the polarized limb is formed by illumination of the inner parts of
%the disk
%much closer to the nucleus than the unpolarised limb, which is seen because
%of the boosting provided by forward unpolarized scattering. 
%
% But what about the forward scattered component from the inner ring? (this
% should be bright and not just concentrated in the nucleus) Also, unless the
% material is distributed in just two rings, which just happen to be at the
% correct inclination to be viewed edge-on, what about the scattered light from
% all the rings at other angles in the warped disk structure?....Clive
%

%
% Another radical suggestion: suppose that the walls of the funnel emit H2 molecular
% emission, with the molecular emission excited by some mechanism other than direct illumination
% by the AGN. Suppose also that the radiation field is anisotropic, with a maximum intensity
% in the direction of the NW-SE limb (see above). Thehe molecules might be dissociated
% along this limb, and there would be no diluting molecular H2 emission, whereas along the
% other limb the molecules could remain intact and give unpolarized intrinsic emission.
% Note that H2 lines of similar strength contribute to both 2.0 and 2.25 bands.
% 

\section{Conclusions and Future Work}

Our NICMOS polarimetry observations of Cygnus~A have demonstrated the existence of
a compact reflection nebula around the hidden core, but one whose polarization
properties are inconsistent with the simplest illumination model suggested by the
imaging data.  
The predominantly axial symmetry of the total intensity imaging is
replaced by axial asymmetry and reflection symmetry about the nucleus in polarized light.
% But it's not quite reflection symmetry.....Clive

We have discussed several mechanisms to explain the near-IR polarization structures. 
While none of these is entirely satisfactory, it is clear that the near-IR polarization
properties have the potential to provide key information about the geometries of
the central emitting regions in AGN, and the near-IR continuum emission mechanism(s).
In this context, it will be interesting in future to make similar 
observations of a large sample of powerful radio galaxies in order to determine whether the
extraordinary IR polarization properties of Cygnus~A are a common feature
of the general population of such objects.  
\newpage\noindent
{\bf Acknowledgments.} 
Based on Observations made with the ESA/NASA {\it Hubble Space Telescope}, 
obtained at the Space Telescope Science Institute, which is operated by the
Association of Universities for Research in Astronomy, Inc., under NASA contract
NAS5-26555. We thank the referee --- Stuart Lumsden --- for useful comments. A. Robinson acknowledges support from the Royal Society.
\vglue 1.0cm\noindent
{\bf References}
\begin{description}
%\reference{arnaud87}Arnaud, K.A., Johnstone, R.M., Fabian, A.C., 
%Crawford, C.S., 
%Nulsen, P.E.J., Shafer, R.A., Mushotsky, R.F., 1987, MNRAS, 227, 241

%\reference{baade54}Baade W., Minkowski R., 1954, ApJ, 119, 215

%\reference{baker97} Baker J.C., 1997, MNRAS, 286, 23

\item Bailey, J.A., Sparks, W.B., Hough, J.H., Axon, D.J., 1986,
Nat, 332, 150

%\reference{barthel89}Barthel P.D., 1989, ApJ, 336, 606

\item Carilli C., Barthel, P.D., 1996, A\&ARev, 7, 1

%Djorgovski S., Weir, N., Matthews, K., Graham, J.R., 1991, ApJ, 372, L67

\item Dreher, J.W., 1979, ApJ, 230, 687

\item Hines, D.C., 1998, NICMOS \& VLT, ESO Workshop and 
Conference Proceedings, 55, Wolfram Freudling and Richard Hook (eds), p63

%\reference{jackson98}Jackson N., Tadhunter, C., Sparks, B., 1998, MNRAS, in %press

\item Jones, T.J., 1989, ApJ, 346, 728

\item Krichbaum, T.P., Alef, W., Witzel, A., 1996, in Cygnus~A --- Study of a 
Radio Galaxy, ed. C.L. Carilli \& D.E., Harris (Cambridge: 
Cambridge University Press), 93.

\item Krist J.E., Hook, R., 1997, TinyTim User Guide, Version 4.4 (Baltimore:STScI)

%\reference{kulkarni98}
%Kulkarni V.P., Calzetti, D., Bergeron, L., Rieke, M., Axon, D., 
%Skinner, C., Colina, L., Sparks, W., Daou, D., Gilmore, D.,
%Holfeltz, S., MacKenty, J., Noll, K., Ritchie, C., Schneider, G., 
%Schultz, A., Storrs, A., Suchkov, A., Thompson, R., 1998, ApJ, 492, L121

%\reference{lawrence91}Lawrence A., 1991, MNRAS, 252, 586

\item Lumsden, S.L., Moore, T.J.T., Smith, C., 
Fujiyoshi, T., Bland-Hawthorn, J., 
Ward, M.J., 1999, MNRAS, 303, 209

\item MacKenty J.W., \etal\, 1997, NICMOS Instrument Handbook, Version 2.0
(Baltimore: STScI)

\item Miyoshi, M., Moran, J., Herrnstein, J., Greenhill, L., 
Nakai, N., Diamond, P., Inoue, M., 1995, Nature, 373, 127

\item 
Ogle P.M., Cohen, M.H., Miller, J.S., Tran, H.D., Fosbury, R.A.E.,
Goodrich, R.W., 1997, ApJ, 482, L37

%\reference{osterbrock76}
%Osterbrock, D.E., Koski, A.T., Phillips. M.M., 1976, ApJ, 206, 898

\item Packham, C., Hough, J.H., Young, S., 
Chrysostomou, A., Bailey, J.A., Axon, D.J., Ward, M.J., 1996,
MNRAS, 278, 406

\item
Packham, C., Young, S., Hough, J.H., Tadhunter, C.N., Axon., 1998,
MNRAS, 297, 939

%\reference{simpson95}Simpson C., 1995, DPhil thesis, University of Oxford

%\reference{simpson94}Simpson C., Ward, M., Kotilainen, J., 1994, MNRAS, 271, 247

%\reference{schreier98}
%Schreier, E.J., Marconi, A., Axon, D.J., Caon, N., Machetto, D., 
%Capetti, A., Hough, J.H., Young, S., Packham, C., 1998, ApJ,  

\item Serkowski, 1958, Acta.Astron., 8, 135

\item Sparks, W.B., Axon, D.J., 1999, PASP, in press
 
\item
Stockton A., Ridgway S.E., Lilly, S., 1994, AJ, 108, 414

\item
Stockton, A., Ridgway, S.E., 1996, in Cygnus~A --- Study of a 
Radio Galaxy, ed. C.L. Carilli \& D.E., Harris (Cambridge: 
Cambridge University Press), 1 

\item
Tadhunter C.N., Scarrott S.M., Rolph C.D., 1990, MNRAS, 246, 163 

%\reference{tadhunter91}Tadhunter C.N., 1991, MNRAS, 251, 46p
   
\item Tadhunter C.N., Metz, S., Robinson, A., 1994, MNRAS, 268, 989 

\item Tadhunter C.N., Packham, C., Axon, D.J., Jackson, N.J.,
Hough, J.H., Robinson, A., Young, S., Sparks, W., 1999, ApJ, 512, L91 

\item
Ueno S., Katsuji K., Minoru N., Yamauchi S., Ward M.J., 1994, ApJ, 431, L1

%\reference{velusamy98}
%Velusamy T., Langer, W.D., 1998, Nat, 392, 685 

\item
Ward M.J., Blanco P.R., Wilson A.S., Nishida M., 1991, ApJ, 382, 115

\item
Ward M.J., 1996, in Cygnus~A --- Study of a Radio Galaxy, ed. C.L. Carilli
\& D.E., Harris (Cambridge: Cambridge University Press), 43

\item van der Marel, R., 1998: 
http://sol.stsci.edu/~marel/software.html

\end{description}
\newpage\noindent
{\bf Figure 1}. Infrared (2.0$\mu$m) polarization images of Cygnus~A. 
Top left -- 
total intensity (Stokes I) at full resolution; top right -- polarization degree at full resolution; bottom left --
polarized intensity at full resolution; and bottom right -- polarization vectors plotted
on a contour map of the polarized intensity image derived from the data smoothed with
a 5$\times$5 pixel box filter, with length of vectors proportional to the percentage polarization.
The line segment in the intensity image shows the direction of the radio axis. At the 
redshift of Cygnus~A, 1.0 arcsecond corresponds to 1.0 kpc for 
$H_0 = 75$ km s$^{-1}$ Mpc$^{-1}$ and $q_0 = 0.0$.

\begin{figure}
\psfig{figure=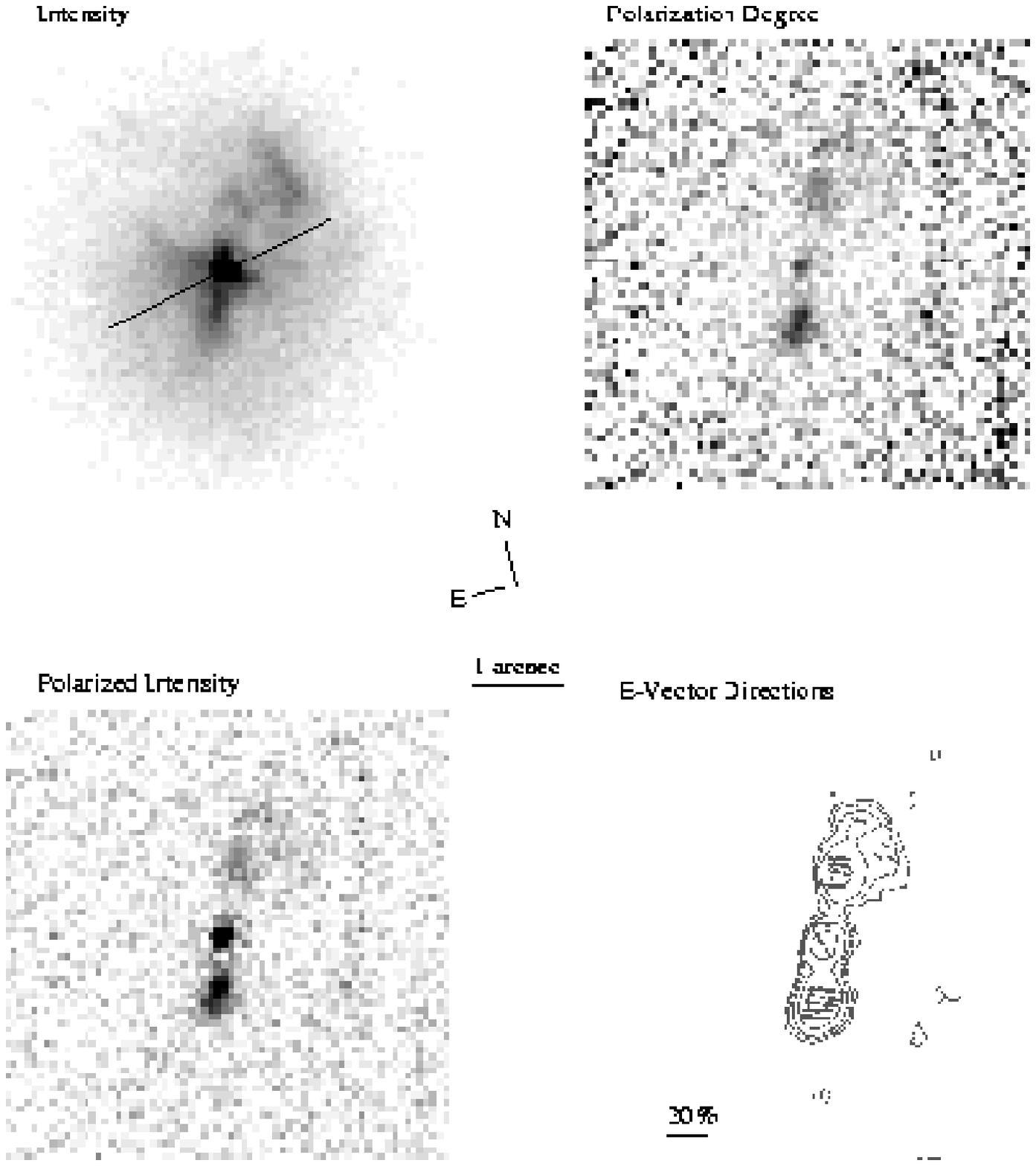,width=15cm}
\end{figure}

\end{document}